\documentclass[conference]{IEEEtran}
\IEEEoverridecommandlockouts

\usepackage{algorithmic}
\usepackage{graphicx}
\usepackage{pgf}
\usepackage{pgfplots}
\usepgfplotslibrary{groupplots}

\usepackage{tikz}
\usetikzlibrary{positioning, shapes, patterns, decorations.pathreplacing,
decorations.markings, calc, fadings, spy, fit, shadows, arrows, fit,
shapes,calc}

\usepackage{xcolor}
\definecolor{sron0}{HTML}{332288}
\definecolor{sron1}{HTML}{88CCEE}
\definecolor{sron2}{HTML}{117733}
\definecolor{sron3}{HTML}{DDCC77}
\definecolor{sron4}{HTML}{CC6677}

\usepackage{amsmath}
\usepackage{amssymb}
\usepackage[hidelinks]{hyperref}
\usepackage{textcomp}
\usepackage{csquotes}
\usepackage{xspace}
\usepackage{caption}
\usepackage{subcaption}

\newcommand*{\ie}{i.e.\@\xspace}

\newtheorem{definition}{Definition}

\usepackage{tabularx}
\usepackage{booktabs}
\usepackage[sort]{cite}

\begin{document}

\title{On the Impact of Attachment Strategies\\for Payment Channel Networks}

\author{
\IEEEauthorblockN{Kimberly Lange}
\IEEEauthorblockA{Distributed Security Infrastructures\\
\textit{Technical University of Berlin}\\
kimberly.lange@campus.tu-berlin.de}
\and
\IEEEauthorblockN{Elias Rohrer}
\IEEEauthorblockA{Distributed Security Infrastructures\\
\textit{Technical University of Berlin}\\
elias.rohrer@tu-berlin.de}
\and
\IEEEauthorblockN{Florian Tschorsch}
\IEEEauthorblockA{Distributed Security Infrastructures\\
\textit{Technical University of Berlin}\\
florian.tschorsch@tu-berlin.de}
}
\maketitle            

\begin{abstract}
    Payment channel networks, such as Bitcoin's Lightning Network, promise to
    improve the scalability of blockchain systems by processing the majority
    of transactions off-chain.
    Due to the design, the positioning of nodes in the network topology
    is a highly influential factor regarding the experienced
    performance, costs, and fee revenue of network participants.
    As a consequence, today's Lightning Network is built around
    a small number of highly-connected hubs.
    Recent literature
    shows the centralizing tendencies to be incentive-compatible and
    at the same time detrimental to security and privacy.
    The choice of attachment strategies therefore becomes a crucial factor
    for the future of such systems.
    In this paper, we provide an empirical study on the (local and global)
    impact of various attachment strategies for
    payment channel networks. To this end, we introduce candidate strategies
    from the field of graph theory and analyze them with respect to their computational
    complexity as well as their repercussions for end users and
    service providers. Moreover, we evaluate their long-term impact on the
    network topology.
\end{abstract}


\section{Introduction}\label{sec:intro}
Payment channel networks, such as Bitcoin's Lightning
Network~\cite{lightning_paper}, are second-layer solutions that aim to improve the scalability,
performance, and privacy aspects of blockchain networks by taking the majority
of transactions off-chain. They allow nodes to open bilateral
payment channels by depositing money in a shared multisig address. By doing this,
parties can negotiate state changes locally in a secure and rapid
fashion.
As each node may establish multiple channels, a network of payment channels
is created, which enables payments to non-adjacent nodes. While such multihop payments
are routed over intermediary nodes, the protocol ensures not only that
the payments are settled atomically, but also that
intermediaries are compensated for their service through transaction fees. In
this way, nodes are incentivized to lock up funds in order to provide the
payment routing infrastructure.

The Lightning Network currently exhibits a high degree
of centralization~\cite{seres2019topology,lin2020centralisation,rohrer2019dischargedPC},
which has shown to be detrimental to the security~\cite{tochner2019hijacking,rohrer2019dischargedPC} and
privacy~\cite{kappos2020empirical,rohrer20countingdown} properties of the network.
Moreover, since the Lightning Network uses a source-routed
best-effort routing protocol to conduct multihop payments, payment reliability is not guaranteed but highly
depends on the connectivity of involved nodes~\cite{waugh2020empirical}. Likewise, the position of
routing nodes in the network topology is highly correlated with their fee revenue.
Therefore, the question arises which connection points are preferable for nodes
joining the network
with respect to their connectivity or revenue.
Prior works studying this question
from a theoretical perspective indicate that profit-optimal join
strategies tend to promote network
centralization~\cite{avarikioti2020ride,sali2020optimizing,ersoy2019profit}.
These results therefore not only suggest the existence of a fundamental
trade-off between the network's goals of efficiency and decentralization, but also a conflict of
interest between the local egoistical point of view of an individual node and
the global long-term development of the network topology.

In this paper, we therefore present an empirical analysis on the local and global
impact of \emph{attachment strategies} for payment channel networks.
We survey the field of graph theory for
strategies that aim to increase the joining node's connectivity and routing
revenue, \ie, we propose strategy candidates from the perspectives of
end-users and service providers. Each strategy is analyzed with
respect to its local impact on the performance of an individual joining node based on network simulations of the
Lightning Network. Moreover, the computational complexities and resource
requirements of all strategies are evaluated in order to classify them according to their real-world
practicability. In contrast, we study the global long-term impact of the
discussed attachment strategies on the Lightning Network topology under the assumption of
mass adoption. To this end, we study how the network's centralization and
performance metrics change in dependence of a given attachment strategy.
We show that, while in the short term centrality-based strategies perform best
in some scenarios, they in the long term result in suboptimal network-wide
transaction success rates and fee costs. To this end, we identify two
candidate strategies with the potential to combine local short-term and global long-term
interests.

The remainder is structured as follows and includes the following
contributions. Section~\ref{sec:prelims}
provides more detailed background information about the Lightning
Network and introduces the model and notations serving as a basis for our research.
In Section~\ref{sec:strategies}, we introduce and discuss attachment strategies
for the Lightning Network, matching different usage scenarios.
The strategies are empirically analyzed from the user and hub perspectives in
Section~\ref{sec:analysis} and their long-term impact is evaluated in
Section~\ref{sec:eval}. After that, Section~\ref{sec:relwork} gives an
overview of related work, before Section~\ref{sec:conclusion} concludes the
paper.

\section{Preliminaries}\label{sec:prelims}
Payment channel networks~(PCNs) establish an overlay network on top of a cryptocurrency.
Instead of storing the details of every transaction on the blockchain,
a PCN offers the possibility to open payment channels
and to process payments bilaterally off-chain.
While different designs of payment channels have been introduced so
far~\cite{hearn2015bitcoin,decker2015duplex,miller2017sprites,decker2018eltoo},
the most popular PCN is the Lightning Network~\cite{lightning_paper}. By the
beginning of May 2020, it consisted of more than 4,300 nodes and around
25,000 payment channels exhibiting a combined capacity of more than 785
bitcoins (more than USD 8 million)\footnote{According to the
network snapshots~\cite{snapshots} provided by~\cite{rohrer2019dischargedPC}.}.
Lightning enables potentially infinite payments between two users with only
two on-chain transactions, the first for opening and the second for closing
the payment channel. The opening transaction allows both parties to securely
deposit money in a shared multisig address on the Bitcoin blockchain. After that, both
users can send bitcoins through the channel by renegotiating the balance allocation
between them. In order to close the channel, its latest
state is published on the blockchain, whereby the final balances are returned to
the involved parties.
Lightning, and PCNs in general, provide mechanisms for resolving conflicts and
attempts of fraud.

\subsection{Network Model}\label{sec:graph_model}
We model the Lightning Network as a \textit{directed multigraph}~$G = (V,E)$,
where the vertex set~$V$ constitutes the Lightning nodes and the multiset~$E$ the payment channels.
Every bidirectional channel is represented by two directed edges in order to
separately store the individual capacities and channel policies of both
channel endpoints. Accordingly, the edge~$(u, v)$ stores how high $u$'s
share of the total channel balance is and which settings $u$ chose for the
channel. As each channel locks funds
and each on-chain transaction involves costly transaction fees,
opening many channels on the Lightning Network can be
expensive. In order to reduce the number of required channels, the Lightning Network offers multihop
routing, which enables the sending of payments to non-adjacent nodes in the
network. In this case, the payment is routed over intermediate nodes along the
payment path, which is determined by the payment's sender and secured by
Hashed Time-Locked Contract~(HTLC) protocols.
For a more detailed description of the HTLC construction the reader is referred to~\cite{mccorry2016towards,lightning_paper}.
Moreover, privacy is improved by applying an onion routing scheme based on the
Sphinx~\cite{danezis2009sphinx} mix packet format.

The payment's sender typically selects the most suitable route by running an adapted version of Dijkstra's shortest path
algorithm~\cite{dijkstra1959note} that considers channel capacities, fees and locking
duration in the edge weight calculation. That is, the algorithm first discards
all candidate edges with insufficient capacities and then selects the path
with minimal aggregated edge weights based on the intermediate nodes' fee
policies and maximum lock-time. Such a weight-based algorithm is for example utilized by the popular \texttt{LND}
implementation, which accounts for more than 90\% of today's network
nodes~\cite{tochner2019hijacking}.
For the calculation of the respective transaction fees, each edge in the public network graph stores the routing fee policies, which are composed of a base fee~$f^B$ and a proportional
fee~$f^P$. The base fee~$f^B$ is a fixed amount that has to be paid to the routing
node for every forwarded payment; the default value is 1\,satoshi ($=1 \cdot 10^{-8}$\,BTC). The
default value of proportional fee~$f^P$ is $1 \cdot 10^{-6}$\,satoshi, which
is multiplied with the transaction amount~$|tx|$ of each payment. Therefore, routing higher
value payments generates higher fees for the routing nodes. Concisely, the
fee~$f_u(v, |tx|)$ that has to be paid to the routing node~$u$ for forwarding a
transaction with amount~$|tx|$ to~$v$ can be calculated accordingly as
\begin{equation*}
f_u(v, |tx|) = f^B_u(v) + f^P_u(v) \cdot |tx|.
\end{equation*}

In order to account for the weight-based routing algorithm, parts of our
graph analysis is based on the \emph{fee graph}~$G_{F,|tx|}$, which we obtain
through a transformation on $G$. This transformation allows the network analysis to account for Lightning's
routing behavior in an approximative fashion, even when applying standard weight-based graph algorithms.
In particular, $G$ is reduced to $G_{F,|tx|}$ by excluding all edges of
insufficient capacities with respect to a transaction amount~$|tx|$. The
weights for each edge~$(u,v)$ in $G_{F,|tx|}$ are set to $f_u(v,|tx|)$, \ie,
they denominate the routing fees that would arise from transferring $|tx|$
through this channel.\footnote{Note that this approximative approach is only applied when necessary
for general graph analysis or as part of the attachment algorithms. In
contrast, the simulation framework used for the evaluation of the proposed
strategies follows a payment protocol that closely resembles the
real-world behavior, as will be discussed in Section~\ref{sub:simulator}.}

\subsection{Joining the Network}
Due to the costs associated with channel establishment, a node joining the
network should follow a certain set of rules for choosing its initial
connection points according to an optimization goal. We call such an algorithm returning a candidate node set $C
\subseteq V$ an \emph{attachment strategy} $\mathcal{S}(G, k, \texttt{cap}) \to C$,
which takes as parameters the public network graph $G$, the number of channels
to be opened $k=|C|$, and the capacity $\texttt{cap}$ (in satoshi) that each
of the channels should hold.

The respective optimization goal depends on the motivation for joining
the network. We consider the attachment strategies from the point-of-view of
three distinct perspectives:
\begin{itemize}
    \item \emph{End-users} join the network to conduct cheap, reliable, and fast
        payments and therefore are interested in strategies that improve their
        local connectivity to the network.
    \item \emph{Service providers} participate as routing nodes in the network in
        order to earn transaction fees. They are therefore interested in
        optimizing their local node's channel selection in order to receive
        maximal profit.
    \item \emph{The network} perspective regards the global impact of a
        particular strategy and considers its impact on the
        network's overall connectivity and reliability over time.
\end{itemize}
As these view points follow partly conflicting interests, they may not easily be
reconciled, but expose a fundamental trade-off between short-term egoistical
efficiency and the long-term development of the network (cf.~\cite{waugh2020empirical}).
However, as different attachment strategies fall on different points in the
spectrum of this trade-off, we empirically investigate their usefulness regarding these
three view points.

The performance of each strategy of course highly depends on the user's
behavior: if we for example assume an end-user
would conduct frequent payments to only a single service provider, the optimal
connectivity-oriented strategy would be to establish a direct payment channel
to it. However, so far no reliable data source on user behavior in payment
channel networks is publicly available to the research community, which
necessitates the introduction of a number of assumptions with regard to the
payment model. To this end, we refrain from introducing overly complex
assumptions that may act as confounding factors to our analysis.
In particular, we assume for the sake of simplicity that the user plans to send payments to destinations all over the
network. Moreover, we assume that the capacity $\texttt{cap}$ is
the same for all $k$ channels and that initial balances are split equally
between the channel endpoints.
We also assume that every node in the
network agrees to open a channel, which may not be the case in the real
network, in particular since recent research found such optimistic behavior to
entail security risks~\cite{harris2020flood}. Finally, we assume new channels
to be established with the default fee settings. 
Note that in current Lightning implementations attachment strategies
are used in the so-called \emph{autopilot} feature that allows the client
software to automatically choose and establish new channels.

\section{Network Attachment Strategies}\label{sec:strategies}
In the following, we introduce candidate strategies for nodes joining payment channel
networks. We also provide a first assessment of their applicability as well as
their complexity in dependence of the number of nodes $n = |V|$
and number of edges $m = |E|$.

\subsection{Random} The \textsf{Random} strategy is the simplest attachment
strategy, in which the attachment points are determined by uniform random
sampling from the node set~$V$. This strategy can be quickly computed in~$\mathcal{O}(n)$
and, while it mainly serves as a
baseline for comparison, it counteracts centralizing tendencies
since it does not prefer any particular connection point.

\subsection{Highest Degree} The \textsf{Highest Degree} strategy sorts all nodes~$V$ according to their degree,
and returns the $k$ nodes with
the highest degrees. As the number of different neighbors is presumably more
meaningful than the total number of channels a node~$v$ has, its degree~$\text{deg}(v)$
is determined in the fee graph~$G_F$, since it disregards
multi-edges. The candidate set can be computed quickly with this strategy
because $\text{deg}(v)$ can be retrieved from the adjacency lists and sorting
can be done in~$\mathcal{O}(n \log n)$.

Connecting to nodes with highest degrees is an extreme form of \emph{preferential
attachment} which is known to induce a \enquote{rich-gets-richer} effect that
yields scale-free networks~\cite{barabasi1999emergence}, and is likely
responsible for the highly centralized substructures found in the Lightning
Network today. In fact, highest-degree attachment strategies were
deployed in prior versions of \texttt{LND}'s autopilot feature and have been
critically discussed in the community~\cite{lnd_autopilot_issue}.

\subsection{Betweenness Centrality}
The notion of \emph{betweenness centrality}~\cite{freeman1977betweenness}
indicates how many shortest paths in the network graph~$G$ a node $v$ is part of. 
More specifically, $bc(v) = \sum_{\substack{s,t \in V\\s \neq v \neq t}}
\frac{\sigma_{st}(v)}{\sigma_{st}}$,
where $\sigma_{st}$ is the total number of shortest paths from~$s$ to~$t$ and $\sigma_{st}(v)$ is the number of shortest paths from~$s$ to~$t$ via~$v$.

In context of Lightning, nodes exhibiting a high betweenness
centrality implies that they are often chosen by the weight-based routing
algorithm and therefore are part of many payment paths.
Since a large share of the network can be reached via these nodes with
minimal distance in terms of fees, they are in return promising candidates for node
attachment. Note that this often also corresponds to overall shorter payment
paths, which improves reliability.

Consequently, the \textsf{Betweenness} attachment strategy elects the $k$ nodes
with the highest betweenness centrality values, which are calculated via the
weighted Brandes' algorithm~\cite{Brandes08} based on the fee graph~$G_F$.
As the weighted version of the algorithm has a runtime complexity in~$\mathcal{O}(nm + n^2\log n)$,
our implementation additionally employs the optimizations from~\cite{fastBC},
which speed up the calculation of betweenness centralities
(but do not change the algorithmic complexity).
Connecting to nodes with the highest betweenness centralities is
another form of preferential attachment and likely results in further network
centralization.

\subsection{$k$-Center} \label{sec:k-center}
The \textsf{$k$-Center} strategy is based on the assumption that the joining
node can improve
its overall connectivity to the network by establishing channels to $k$ nodes such that
the highest distances between them and any other node in the network are
minimized. Ideally, this would lead to nodes in different parts of the network being chosen as the $k$ new neighbors in order to minimize the length of the longest shortest payment
path. This likely results in faster and cheaper transactions due to fewer
nodes being part of the routes. Reducing the number of nodes and channels
contained in a payment route can also decrease the risk that a transaction
fails as there are less points of failure.

The idea for this strategy is based on the $k$-center problem \cite{hochbaum1985k_center},
which is defined as follows.
\begin{definition}
Given a complete undirected graph~$G = (V, E)$ in a metric space and an integer~$k$, a $k$-center is a subset of nodes~$C \subseteq V$ with~$|C| \leq k$ such that~$max_{v \in V} d(v,C)$ is minimized, with~$d(v,C)$ being the shortest distance of~$v$ to the closest node in~$C$.
\end{definition}
It was previously proven that this problem is NP-complete and that it is NP-hard
even for an $\epsilon$\hbox{-}approximation with~$\epsilon <
2$~\cite{hochbaum1985k_center}. This means that 2-approximation algorithms,
which return a solution that is within twice the optimal solution value in
polynomial time, are the best possible algorithms for the $k$-center problem,
unless~$P = NP$. Due to the fact that distances in the fee graph~$G_F$ are not necessarily
symmetric, the Lightning Network unfortunately cannot be modeled as a
weighted fee graph in metric space. We therefore use the greedy $k$-center algorithm
introduced in~\cite{gonzalez} on a generated complete \emph{distance graph} to minimize the
number of hops on the longest shortest path and disregard fees or channel
capacities. To this end, the joining node first establishes a
connection to the network's highest degree node and then executes a single-source shortest path~(SSSP)
search to retrieve the distances for the $k$-center algorithm. This results in
a total time complexity of~$\mathcal{O}(k(m+n))$.

As the \textsf{$k$-Center} strategy aims to interconnect the network centers, it should improve the network's robustness and facilitate decentralization.

\subsection{$k$-Median} \label{sec:k-median}
Besides looking at the longest shortest path to any other node in the network,
a promising strategy is to minimize the average shortest path distance to all
other nodes. Assuming that the joining node sends a transaction to any other node with the
same probability, it is very likely favorable to require a minimal average
number of hops to any other node in order to reduce transaction fees,
latencies, and failures. Hence, we have to solve a problem that is known as the \textit{Single-Source Average Shortest Path Distance
Minimization}~(SS-ASPDM) problem~\cite{meyerson2009min_avgSP}. It was previously proven that an optimal solution to the SS\hbox{-}ASPDM problem
for a node~$v$ can be found by only adding edges incident
to~$v$~\cite{meyerson2009min_avgSP}. Thus, the problem can be utilized in our use
case of the Lightning Network since a joining node may only influence the opening of
channels which are incident to itself. Adopting the approach to only add
edges incident to the source node~$v$, the
SS\hbox{-}ASPDM problem corresponds to the $k$-median problem~\cite{meyerson2009min_avgSP}.

In a graph context, the $k$-median problem can be formulated as follows.
\begin{definition}
Given a complete undirected graph~$G = (V, E)$ in a metric space and an
integer~$k$, the $k$-median problem strives to find a subset of nodes~$C \subseteq V$ with $|C| \leq k$ such that~$\sum_{v \in V} d(v,C)$ is minimized, with~$d(v,C)$ being the shortest distance of~$v$ to the nearest node in~$C$.
\end{definition}
Again, the problem is NP-hard~\cite{chrobak2005RGreedy} and only an
approximate solution can be found within polynomial time, unless~$P = NP$.
For solving the $k$-median problem in a distance graph, we establish an
initial connection to the highest degree node and then utilize the \enquote{forward}
greedy algorithm presented in~\cite{chrobak2005RGreedy}, which results
in an overall time complexity of~$\mathcal{O}(kn(n+m)\log n)$ when applied to
the weighted fee graph.

Similarly to the \textsf{$k$-Center} approach, the \textsf{$k$-Median}
strategy promises to improve network robustness and reduce centralization.

\subsection{Maximum Betweenness Improvement (MBI)}
A node that joins the network with the intent to act as a service provider or
routing node strives for financial profit from participating in the Lightning Network.
To this end, a routing node $v$ should rather focus on optimizing its own betweenness centrality $bc(v)$
than connecting to central nodes.

Therefore, it has to solve a problem known as Maximum Betweenness
Improvement~(MBI)~\cite{bergamini2018betweenness}, which is defined as follows.
\begin{definition}[MBI]\label{def:mbi}
Given a directed graph $G$, a node $v$, and an integer $k$, which set of edges $S$ incident to $v$, with $|S| \leq k$, should be added to $G$ in order to maximize $bc(v)$?
\end{definition}
The MBI problem has been proven to be
NP-hard, but a greedy algorithm that provides an approximate solution
exists~\cite{bergamini2018betweenness}.

This \textsf{MBI} strategy temporarily opens any channel that node $v$
could set up, calculates $bc(v)$, and closes the channel again. This is
repeated for all possible channels and in the end the channel generating the
highest betweenness improvement for the joining node is elected. This channel is then
established and the procedure is repeated until all $k$ candidates
are found, leading to an overall high time complexity in~$\mathcal{O}(kn^3)$.

Note that this strategy is similar to the approach found
in~\cite{ersoy2019profit}, which however also optimizes
the node's fee settings. As this results in a further increased
computational complexity over the already high resource requirements of
Bergamini et al.'s algorithm, we in lieu of these optimizations follow the more feasible \textsf{MBI}
strategy.

\section{Empirical Analysis}\label{sec:analysis}
\begin{figure*}[t]
	\centering
	\resizebox{0.9\linewidth}{!}{
	\input{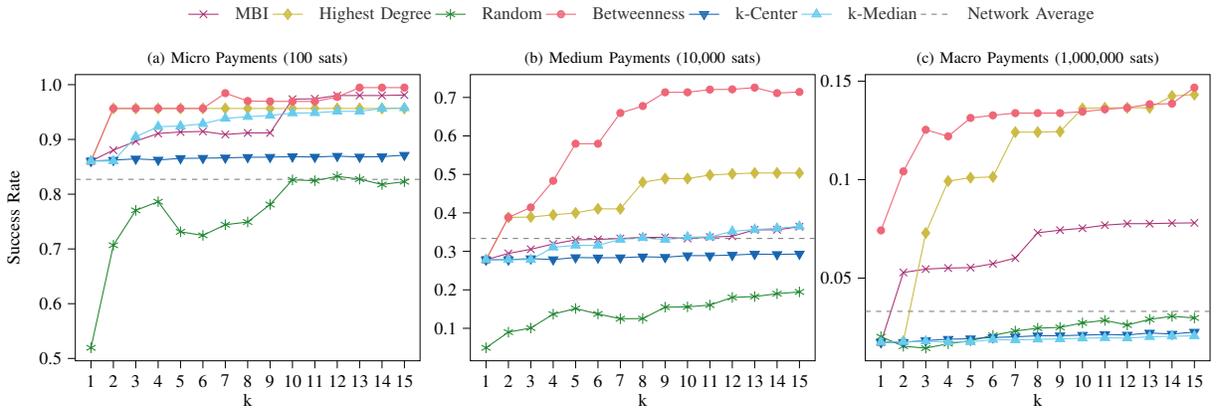}
    }
	\caption{Transaction success rates in dependence of chosen
    attachment strategy and transaction amounts.}
    \label{fig:success}
\end{figure*}
In the following, we empirically analyze the performance of attachment strategies for
payment channel networks from a local perspective, \ie, from the view of a
single end-user or service provider aiming to join the network.

\subsection{Network Simulator, Setup, and Methodology}
\label{sub:simulator}
As a basis for the empirical analysis, we developed a time-discrete event simulator that
implements the network multigraph model (cf.~Section~\ref{sec:graph_model})
and allows to simulate payment processing as well as nodes joining the network according
to a given attachment strategy.\footnote{The simulator code base is publicly
available in our companion repository at \url{https://gitlab.tu-berlin.de/rohrer/pcn-attachment-data}.} 
The simulator initially reads the network graph from a snapshot of the
Lightning Network and simulates path finding through a weight-based route selection
algorithm similar to the one found in \texttt{LND}. 
While some aspects of the real-world payment procedure---such as
the HTLC protocol negotiations---are omitted by our simulation model for the sake of
simplicity, the simulator was carefully implemented to approximate the real-world
behavior.
To this end, transaction processing
is simulated by checking and adjusting the available balances along the
payment path. During this phase, the arising fee revenues are calculated based
on the provided fee policies and the remaining transaction value for each hop
along the way.
Note that consequentially and just as in the real network,  transaction success is not guaranteed even if a path is
found, as the path finding algorithm does not operate on the private balances, but the public capacities.
We base our further analysis on a snapshot of the Lightning Network from
May~1, 2020 at 10am that was taken from the dataset~\cite{snapshots} provided
by~\cite{rohrer2019dischargedPC}.
At this time, the largest connected component of the network consisted of more than 4,300~nodes connected by nearly 25,000~channels, which
held an overall capacity of more than 785~BTC.

In order to analyze their performance, we simulated the joining of individual
nodes according to the given attachment strategy, every time establishing \mbox{$k\in\{1,\ldots,15\}$} channels
with sufficient capacity and default fee settings. We then evaluated the
connectivity and fee revenue of the joined node through two sets of simulated
payments: one set of 1,000 transactions with the joined node as a fixed source and
the destination selected by uniform random sampling, and another set of 1,000~transactions
for which both source and destination were chosen randomly.
The simulations were conducted
under the assumption of three different transaction volumes: micro payments of
100~sats, medium payments of 10,000~sats, and macro payments of 1,000,000~sats
(see also~\cite{ersoy2019profit}). If not stated otherwise,
our analysis is based on the most relaxed assumption of 100~sats.
For every strategy, transaction value, and every value of~$k$, the simulations
were furthermore repeated 30 times with different seed value inputs for the
utilized random number generator. This results in
a five-digit sample size ensuring the statistical
significance of the results.

\subsection{Transaction Success}
In order to assess the impact of the attachment strategies on the
connectivity of the joining node, we analyzed the average transaction success
rate, \ie, the share of all transaction that actually succeeded. In
Figure~\ref{fig:success}, the average success rate is shown in
dependence of the number of transaction amounts and channels $k$ that were
established corresponding to the respective strategies. Moreover, the a priori network-wide average
success rate is shown for comparison, which was determined by simulating
10,000 transactions with randomly chosen sources and destinations in the
initial graph configuration.

We observe that generally node connectivity improves with the number
of established channels and that all but the \textsf{Random} strategy
tend to result in an average success rate higher than the network average. Moreover,
strategies that prefer central connection points, such as the
\textsf{Betweenness} strategy, fare better than strategies that connect the
periphery of the network, such as \textsf{$k$-Center}. This is likely the case
because connecting to very central points in the network reduces the average
path length and thereby also the probability of routing failures due to
unavailable balances.

This is supported by the fact that the assumed transaction volume has a big impact on
the average success rate: while the network average for micro payments is around 83\% (Figure~\ref{fig:successa}), it drops below 34\% for medium payments
(Figure~\ref{fig:successb}), and even to less
than 4\% for macro payments (Figure~\ref{fig:successc}). This observation is of course in line with
previous literature, in which Lightning's limited available capacity and
the resulting low success rates for higher-volume payments have been discussed
for some time~\cite{waugh2020empirical,rohrer2019dischargedPC,beres2019simulator}.
Our results underline that currently only a small number of central nodes hold
enough capacity to be able to route any high-volume payments.
While the heavily skewed capacity distribution results in overall very
low transaction success rates, we observe that
strategies that preferably connect to these few central nodes---such as
\textsf{Betweenness}, \textsf{Highest Degree}, and \textsf{MBI}---can increase
their lead in such high-volume payment scenarios.
However, in order to limit the impact the current
capacity constraints found in the Lightning Network have on our results, we continue our
further analysis of attachment strategies under the most relaxed assumption of micro payments.

\subsection{Transaction Fees}
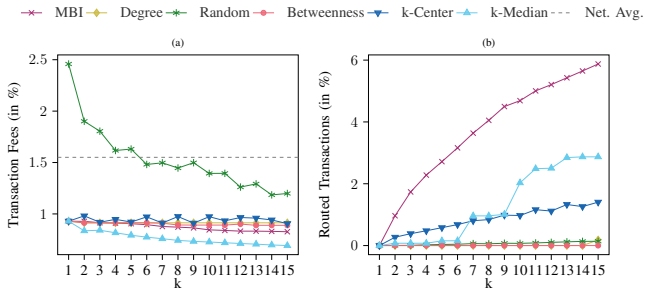
\begin{figure}[t]
	\centering
    \resizebox{\columnwidth}{!}{
    \large
	\begin{tikzpicture}

\definecolor{color0}{rgb}{0.666666666666667,0.2,0.466666666666667}
\definecolor{color1}{rgb}{0.8,0.733333333333333,0.266666666666667}
\definecolor{color2}{rgb}{0.133333333333333,0.533333333333333,0.2}
\definecolor{color3}{rgb}{0.933333333333333,0.4,0.466666666666667}
\definecolor{color4}{rgb}{0.0980392156862745,0.396078431372549,0.690196078431373}
\definecolor{color5}{rgb}{0.4,0.8,0.933333333333333}

\begin{groupplot}[group style={group size=2 by 1, group name=whatever, horizontal sep=2cm}]
\nextgroupplot[
legend style={fill opacity=0.8, draw opacity=1, text opacity=1, draw=white!80!black},
legend to name=named,
legend style={draw=none},
legend columns=7,
tick align=outside,
tick pos=left,
x grid style={white!69.0196078431373!black},
xlabel={k},
xmin=-0.7, xmax=14.7,
xtick style={color=black},
xtick={0,1,2,3,4,5,6,7,8,9,10,11,12,13,14},
xtick={0,1,2,3,4,5,6,7,8,9,10,11,12,13,14},
xtick={0,1,2,3,4,5,6,7,8,9,10,11,12,13,14},
xtick={0,1,2,3,4,5,6,7,8,9,10,11,12,13,14},
xtick={0,1,2,3,4,5,6,7,8,9,10,11,12,13,14},
xtick={0,1,2,3,4,5,6,7,8,9,10,11,12,13,14},
xtick={0,1,2,3,4,5,6,7,8,9,10,11,12,13,14},
xticklabels={1,2,3,4,5,6,7,8,9,10,11,12,13,14,15},
xticklabels={1,2,3,4,5,6,7,8,9,10,11,12,13,14,15},
xticklabels={1,2,3,4,5,6,7,8,9,10,11,12,13,14,15},
xticklabels={1,2,3,4,5,6,7,8,9,10,11,12,13,14,15},
xticklabels={1,2,3,4,5,6,7,8,9,10,11,12,13,14,15},
xticklabels={1,2,3,4,5,6,7,8,9,10,11,12,13,14,15},
xticklabels={1,2,3,4,5,6,7,8,9,10,11,12,13,14,15},
y grid style={white!69.0196078431373!black},
ylabel={Transaction Fees (in \%)},
ymin=0.603946374619541, ymax=2.54665206980724,
ytick style={color=black}
]
\addplot [semithick, color0, mark=x, mark size=3, mark options={solid}]
table {%
0 0.929283465689889
1 0.927556501399866
2 0.914419859080423
3 0.910622726398094
4 0.905009537863663
5 0.895517439145995
6 0.877666293822206
7 0.871318694419058
8 0.864374135308003
9 0.844980499086632
10 0.839430809747143
11 0.832453026821461
12 0.832453026821461
13 0.830311533452119
14 0.826829843576382
};
\addlegendentry{MBI}
\addplot [semithick, color1, mark=diamond*, mark size=3, mark options={solid}]
table {%
0 0.929283465689889
1 0.91349608660645
2 0.913496086606451
3 0.913496086606451
4 0.913496086606451
5 0.9134955345463
6 0.9134955345463
7 0.9134955345463
8 0.9134955345463
9 0.9134955345463
10 0.913494309110723
11 0.913494309110723
12 0.913494309110723
13 0.913494309110723
14 0.913494309110723
};
\addlegendentry{Degree}
\addplot [semithick, color2, mark=asterisk, mark size=3, mark options={solid}]
table {%
0 2.45834726548053
1 1.90113927656964
2 1.80477280850374
3 1.6168616308378
4 1.63103705543153
5 1.48117513146615
6 1.49702633905232
7 1.44651395916379
8 1.49825462684948
9 1.39399856386583
10 1.39483279967593
11 1.26123310718571
12 1.29210700060147
13 1.18517613175772
14 1.19870512477747
};
\addlegendentry{Random}
\addplot [semithick, color3, mark=*, mark size=2, mark options={solid}]
table {%
0 0.929283465689889
1 0.913496086606451
2 0.913496086606451
3 0.913496086606451
4 0.913496086606451
5 0.913496086606451
6 0.910013905779136
7 0.89226650266036
8 0.891505364011421
9 0.891505364011421
10 0.891505364011421
11 0.901266431406555
12 0.889237822312714
13 0.889237822312714
14 0.889237822312714
};
\addlegendentry{Betweenness}
\addplot [semithick, color4, mark=triangle*, mark size=3, mark options={solid,rotate=180}]
table {%
0 0.929283465689889
1 0.980857476289273
2 0.918910420577949
3 0.947248816048996
4 0.921005344089499
5 0.971429432430817
6 0.911509103789522
7 0.974613918102023
8 0.91042016273318
9 0.973083851068513
10 0.933119232244864
11 0.964936724326717
12 0.959317274919864
13 0.941377387271595
14 0.905782484099708
};
\addlegendentry{k-Center}
\addplot [semithick, color5, mark=triangle*, mark size=3, mark options={solid}]
table {%
0 0.929283465689889
1 0.835972220432632
2 0.841370110686628
3 0.816635634500521
4 0.79275660180557
5 0.774124991464213
6 0.758803520540087
7 0.74275876371892
8 0.73306788754411
9 0.726704478205154
10 0.71941594759857
11 0.712363115046871
12 0.705861881677566
13 0.698148474806334
14 0.692251178946255
};
\addlegendentry{k-Median}
\addplot [semithick, white!50.1960784313725!black, dashed]
table {%
-0.699999999999999 1.55046919789329
14.7 1.55046919789329
};
\addlegendentry{Net. Avg.}

\nextgroupplot[
tick align=outside,
tick pos=left,
x grid style={white!69.0196078431373!black},
xlabel={k},
xmin=-0.7, xmax=14.7,
xtick style={color=black},
xtick={0,1,2,3,4,5,6,7,8,9,10,11,12,13,14},
xtick={0,1,2,3,4,5,6,7,8,9,10,11,12,13,14},
xtick={0,1,2,3,4,5,6,7,8,9,10,11,12,13,14},
xtick={0,1,2,3,4,5,6,7,8,9,10,11,12,13,14},
xtick={0,1,2,3,4,5,6,7,8,9,10,11,12,13,14},
xtick={0,1,2,3,4,5,6,7,8,9,10,11,12,13,14},
xtick={0,1,2,3,4,5,6,7,8,9,10,11,12,13,14},
xticklabels={1,2,3,4,5,6,7,8,9,10,11,12,13,14,15},
xticklabels={1,2,3,4,5,6,7,8,9,10,11,12,13,14,15},
xticklabels={1,2,3,4,5,6,7,8,9,10,11,12,13,14,15},
xticklabels={1,2,3,4,5,6,7,8,9,10,11,12,13,14,15},
xticklabels={1,2,3,4,5,6,7,8,9,10,11,12,13,14,15},
xticklabels={1,2,3,4,5,6,7,8,9,10,11,12,13,14,15},
xticklabels={1,2,3,4,5,6,7,8,9,10,11,12,13,14,15},
y grid style={white!69.0196078431373!black},
ylabel={Routed Transactions (in \%)},
ymin=-0.293719639139486, ymax=6.16811242192922,
ytick style={color=black}
]
\addplot [semithick, color0, mark=x, mark size=3, mark options={solid}]
table {%
0 0
1 0.964313366650243
2 1.73497699575036
3 2.27917235584497
4 2.71490046531155
5 3.16208292614826
6 3.63674365137423
7 4.05141782205811
8 4.49763099219621
9 4.68793535803845
10 5.00503629606474
11 5.20703070723913
12 5.42877982702928
13 5.6486589639499
14 5.87439278278973
};
\addplot [semithick, color1, mark=diamond*, mark size=3, mark options={solid}]
table {%
0 0
1 0
2 0
3 0
4 0
5 0
6 0
7 0
8 0
9 0
10 0
11 0
12 0
13 0
14 0.181236673773987
};
\addplot [semithick, color2, mark=asterisk, mark size=3, mark options={solid}]
table {%
0 0
1 0.00355378655957923
2 0.0070977358222727
3 0.0248835803917387
4 0.0354874197097129
5 0.0426575663858377
6 0.0638818894843312
7 0.0568828213879408
8 0.0745288710650531
9 0.0711009989690355
10 0.0851818988464951
11 0.103081790068603
12 0.120665791248181
13 0.135082293555153
14 0.138425498686732
};
\addplot [semithick, color3, mark=*, mark size=2, mark options={solid}]
table {%
0 0
1 0
2 0
3 0
4 0
5 0
6 0
7 0
8 0
9 0
10 0
11 0
12 0
13 0
14 0
};
\addplot [semithick, color4, mark=triangle*, mark size=3, mark options={solid,rotate=180}]
table {%
0 0
1 0.269972647508081
2 0.384150245429323
3 0.478553704360156
4 0.579431943407629
5 0.677833771027042
6 0.803298500035544
7 0.84036593149422
8 0.983664772727273
9 0.971975877970912
10 1.16047980694159
11 1.11646700219749
12 1.32807783814495
13 1.26074295049364
14 1.40480329206428
};
\addplot [semithick, color5, mark=triangle*, mark size=3, mark options={solid}]
table {%
0 0
1 0.0746268656716418
2 0.0746242137806048
3 0.0746242137806048
4 0.152801961550762
5 0.156355495540315
6 0.960788484719563
7 0.960788484719563
8 1.01038749246641
9 2.02992203413641
10 2.48824313890644
11 2.50228118200323
12 2.84621323787464
13 2.87047759413272
14 2.87047759413272
};
\end{groupplot}

\node[text width=6cm,align=center,anchor=south] at (whatever c1r1.north)
{\subcaption{\label{fig:fees}}};
\node[text width=6cm,align=center,anchor=south] at (whatever c2r1.north)
{\subcaption{\label{fig:routing}}};
\path ([yshift=+5mm]whatever c1r1.north) -- node[shift={(0,.7)}] {\ref{named}} ([yshift=+5mm]whatever c2r1.north);

\end{tikzpicture}
    }
    \caption{Transaction fees and share of routed transactions in dependence of
    chosen attachment strategy.}
    \label{fig:feesrouting}
\end{figure}
\begin{table*}[t]
    \centering
    \caption{Algorithm runtimes in dependence of chosen attachment strategy and
    number $k$ of established channels (in sec.).}
    \label{tab:runtimes}
    \begin{tabularx}{\linewidth}{Xrrrrrrrrrrrrrrr}
    \toprule
    & 1 & 2 & 3 & 4 & 5 & 6 & 7 & 8 & 9 & 10\\
    \midrule
    \textsf{Highest Degree}& 0.25&0.25&0.25&0.25&0.25&0.25&0.25&0.25&0.25&0.25\\
    \textsf{Betweenness}&440.00&440.00&440.00&440.00&440.00&440.00&440.00&440.00&440.00&440.00\\
    \textsf{$k$-Median}&2.70&4.70&6.40&8.50&9.70&11.30&13.30&15.00&17.10&18.20\\
    \textsf{$k$-Center}&0.51&0.59&0.63&0.66&0.67&0.72&0.75&0.81&0.88&0.81\\
    \textsf{MBI}&2,784.00&4,834.00&6,965.00&9,232.00&11,429.00&13,938.00&16,478.00&19,371.00&22,294.00&24,634.00\\
    \bottomrule
    \end{tabularx}
\end{table*}

End-users joining the Lightning Network likely want to optimize their
connection point with regards to the result fees that arise from sending
payments. In Figure~\ref{fig:fees} the fees paid by the
connecting node are shown in dependence of the number of channels and with respect
to the chosen strategy. Again, generally all strategies result in fee
costs lower than the network average, which is even true for \textsf{Random}
for more than $k=5$ channels. As the fees in most cases improve linearly with the
number of established channels, it can be concluded that overall better
connectivity and the resulting increased routing opportunities help to reduce
the cost associated with sending Lightning payments.

Interestingly, the \textsf{$k$-Median} strategy is the clear favorite with
regards to fee saving, likely as it helps to increase connectivity between
network clusters and connects these as well as more centralized nodes.

\subsection{Service Provider Revenue}
Service providers join the network with the intend to earn the maximum amount
of profit. To this end, we analyze which attachment strategy can help to
improve their fee revenue.
The share of routed transactions (which in our case directly corresponds to
the fee revenue) is shown in Figure~\ref{fig:routing}.\footnote{Note that our analysis
compares the proportional fee revenues gained from routing in the Lightning Network and
does not consider any costs for running a routing node, such as
the on-chain fees associated with channel establishment. In order to estimate
the net.\ profit of a node operator, such cost would have to be known and
subtracted from the revenue.}

Independently of the strategy, the share of routed transactions improves with the overall
connectivity of the joining nodes, \ie, it increases with the number of
established payment channels $k$, but tends to favor strategies that improve
path diversity. However, the \textsf{MBI} strategy is clearly
superior in this regard, allowing the joining node even to route close to
6\% of all payments conducted in the network by establishing $k=15$ channels.
This comes to no surprise as this strategy is specifically focused on
maximizing the number of payment paths routed through the joining node, and
previous work showed the benefits of such an approach~\cite{ersoy2019profit}.
Apart from this, the \textsf{$k$-Median} strategy is a promising candidate, as
it able to secure the service provider a routing share of close to 3\% of all
payments in the case of $k=15$.

\subsection{Runtime Analysis}\label{sec:runtimes}

In order for a attachment strategy to be an actual candidate to be implemented
in the autopilot functionality of a Lightning client implementation, it should deliver its
results in a viable amount of time. Therefore, we measured the
run times of discussed strategies under real-world conditions. To this end, we
deployed our strategy implementations on an \texttt{t2.xlarge}
instance (4~vCPUs based on Intel Xeon 3.3~GHz, 16~GB memory) on Amazon Elastic Compute Cloud~(EC2) running Ubuntu
Server~18.04. We then measured the execution time that it took the algorithms
to return the respective candidate sets.

The results shown in Table~\ref{tab:runtimes} generally concur with our
complexity analysis given in Section~\ref{sec:strategies}: while the
\textsf{Highest Degree}, \textsf{Betweenness} and \textsf{$k$-Center}
strategies remain roughly constant runtimes, \textsf{$k$-Median} and especially \textsf{MBI} grow in
a linear fashion with the number of established payment channels $k$.

This is of particular significance, since it takes \textsf{MBI} between 2,000
and 2,500 seconds longer to finish for each additional channel. As this amounts
to an overall runtime of around seven hours for $k=10$, the practicability of
this strategy is heavily put under question, potentially even given its performance
benefits in terms of fee revenue.

\section{Evaluating the Long-term Impact}\label{sec:eval}
\begin{figure*}[t]
	\centering
	\resizebox{\linewidth}{!}{
    \large
\begin{tikzpicture}

\definecolor{color0}{rgb}{0.8,0.733333333333333,0.266666666666667}
\definecolor{color1}{rgb}{0.133333333333333,0.533333333333333,0.2}
\definecolor{color2}{rgb}{0.933333333333333,0.4,0.466666666666667}
\definecolor{color3}{rgb}{0.0980392156862745,0.396078431372549,0.690196078431373}
\definecolor{color4}{rgb}{0.4,0.8,0.933333333333333}

\begin{groupplot}[group style={group size=4 by 1, group name=whatever, horizontal sep=2cm}]
\nextgroupplot[
legend style={fill opacity=0.8, draw opacity=1, text opacity=1, draw=white!80!black},
legend to name=named,
legend style={draw=none},
legend columns=5,
tick align=outside,
tick pos=left,
x grid style={white!69.0196078431373!black},
xlabel={Nodes Added},
xmin=-250, xmax=5250,
xtick style={color=black},
y grid style={white!69.0196078431373!black},
ylabel={Gini Coefficient (Degree)},
ymin=0.26425230666194, ymax=0.768001130876819,
ytick style={color=black},
ytick={0.2,0.3,0.4,0.5,0.6,0.7,0.8},
yticklabels={0.2,0.3,0.4,0.5,0.6,0.7,0.8}
]
\addplot [semithick, color0, mark=diamond*, mark size=3, mark options={solid}]
table {%
0 0.74510345704887
500 0.735423375478355
1000 0.723104635281673
1500 0.710479225916265
2000 0.698416898914881
2500 0.687218110561251
3000 0.676949187992179
3500 0.667579591582498
4000 0.659041302426626
4500 0.65125522124086
5000 0.644142849677109
};
\addlegendentry{Highest Degree}
\addplot [semithick, color1, mark=asterisk, mark size=3, mark options={solid}]
table {%
0 0.74510345704887
500 0.641415207591523
1000 0.561144548581484
1500 0.497915395349086
2000 0.447408450890006
2500 0.406492555186491
3000 0.372990573830236
3500 0.345501616057231
4000 0.322467362134867
4500 0.303288532881133
5000 0.287149980489889
};
\addlegendentry{Random}
\addplot [semithick, color2, mark=*, mark size=2, mark options={solid}]
table {%
0 0.74510345704887
500 0.73485624107766
1000 0.72264975053734
1500 0.710118068945872
2000 0.698123294486144
2500 0.686974737357741
3000 0.676744103877251
3500 0.667404346603237
4000 0.6588897747363
4500 0.651122865509076
5000 0.644026218060242
};
\addlegendentry{Betweenness}
\addplot [semithick, color3, mark=triangle*, mark size=3, mark options={solid,rotate=180}]
table {%
0 0.74510345704887
500 0.633898737212634
1000 0.546774777627816
1500 0.476119166712978
2000 0.419014306491036
2500 0.373251569763645
3000 0.338606757819956
3500 0.318645123759805
4000 0.313199574797838
4500 0.312849574244384
5000 0.314191167086807
};
\addlegendentry{k-Center}
\addplot [semithick, color4, mark=triangle*, mark size=3, mark options={solid}]
table {%
0 0.74510345704887
500 0.734708941872706
1000 0.722540507557654
1500 0.710019304059377
2000 0.698032475736204
2500 0.686890405675207
3000 0.676665341209406
3500 0.667330468477073
};
\addlegendentry{k-Median}

\nextgroupplot[
tick align=outside,
tick pos=left,
x grid style={white!69.0196078431373!black},
xlabel={Nodes Added},
xmin=-250, xmax=5250,
xtick style={color=black},
y grid style={white!69.0196078431373!black},
ylabel={Diameter},
ymin=3.55, ymax=13.45,
ytick style={color=black}
]
\addplot [semithick, color0, mark=diamond*, mark size=3, mark options={solid}]
table {%
0 13
500 13
1000 13
1500 13
2000 13
2500 13
3000 13
3500 13
4000 13
4500 13
5000 13
};
\addplot [semithick, color1, mark=asterisk, mark size=3, mark options={solid}]
table {%
0 13
500 8.86666666666667
1000 7.9
1500 7.1
2000 6.73333333333333
2500 6.43333333333333
3000 6.2
3500 6.06666666666667
4000 6.03333333333333
4500 6
5000 6
};
\addplot [semithick, color2, mark=*, mark size=2, mark options={solid}]
table {%
0 13
500 13
1000 13
1500 13
2000 13
2500 13
3000 13
3500 13
4000 13
4500 13
5000 13
};
\addplot [semithick, color3, mark=triangle*, mark size=3, mark options={solid,rotate=180}]
table {%
0 13
500 4
1000 4
1500 4
2000 4
2500 4
3000 4
3500 4
4000 4
4500 4
5000 4
};
\addplot [semithick, color4, mark=triangle*, mark size=3, mark options={solid}]
table {%
0 13
500 13
1000 13
1500 13
2000 13
2500 13
3000 13
3500 13
};

\nextgroupplot[
legend style={fill opacity=0.8, draw opacity=1, text opacity=1, draw=white!80!black},
legend to name=named,
legend style={draw=none},
legend columns=5,
tick align=outside,
tick pos=left,
x grid style={white!69.0196078431373!black},
xlabel={Nodes Added},
xmin=-250, xmax=5250,
xtick style={color=black},
y grid style={white!69.0196078431373!black},
ylabel={Success Rate (in \%)},
ymin=82.5554666666667, ymax=100.8272,
ytick style={color=black}
]
\addplot [semithick, color0, mark=diamond*, mark size=3, mark options={solid}]
table {%
0 83.386
500 94.0433333333333
1000 94.4033333333334
1500 94.77
2000 95.2266666666667
2500 95.8533333333333
3000 96.12
3500 96.5966666666666
4000 96.58
4500 96.8666666666667
5000 96.98
};
\addlegendentry{Highest Degree}
\addplot [semithick, color1, mark=asterisk, mark size=3, mark options={solid}]
table {%
0 83.386
500 97.52
1000 98.82
1500 99.37
2000 99.4333333333333
2500 99.7766666666666
3000 99.83
3500 99.89
4000 99.8966666666667
4500 99.95
5000 99.9633333333333
};
\addlegendentry{Random}
\addplot [semithick, color2, mark=*, mark size=2, mark options={solid}]
table {%
0 83.386
500 95.3433333333333
1000 96.2566666666667
1500 97.05
2000 97.96
2500 98.4266666666667
3000 98.8066666666667
3500 99.0366666666667
4000 99.2633333333333
4500 99.44
5000 99.6166666666667
};
\addlegendentry{Betweenness}
\addplot [semithick, color3, mark=triangle*, mark size=3, mark options={solid,rotate=180}]
table {%
0 83.386
500 99.8066666666667
1000 99.94
1500 99.97
2000 99.98
2500 99.9866666666667
3000 99.9533333333333
3500 99.9966666666667
4000 99.99
4500 99.9966666666667
5000 99.9966666666667
};
\addlegendentry{k-Center}
\addplot [semithick, color4, mark=triangle*, mark size=3, mark options={solid}]
table {%
0 83.386
500 96.23
1000 97.1633333333333
1500 97.79
2000 98.4166666666667
2500 98.9166666666667
3000 99.2566666666667
3500 99.4333333333333
};
\addlegendentry{k-Median}

\nextgroupplot[
tick align=outside,
tick pos=left,
x grid style={white!69.0196078431373!black},
xlabel={Nodes Added},
xmin=-250, xmax=5250,
xtick style={color=black},
y grid style={white!69.0196078431373!black},
ylabel={Fees (in \%)},
ymin=1.06224824926854, ymax=1.50751164137468,
ytick style={color=black},
ytick={1.,1.1,1.2,1.3,1.4,1.5,1.6},
yticklabels={1.0,1.1,1.2,1.3,1.4,1.5,1.6}
]
\addplot [semithick, color0, mark=diamond*, mark size=3, mark options={solid}]
table {%
0 1.47243335093012
500 1.46340080566732
1000 1.33348450986286
1500 1.29031934956392
2000 1.32889398187424
2500 1.35525558493848
3000 1.31645737436145
3500 1.22997040263654
4000 1.29376008493927
4500 1.21478397143492
5000 1.24424838013372
};
\addplot [semithick, color1, mark=asterisk, mark size=3, mark options={solid}]
table {%
0 1.47243335093012
500 1.30490889755528
1000 1.31751331208132
1500 1.2849236459757
2000 1.30458009746025
2500 1.33771366378493
3000 1.33826865860087
3500 1.35475493308405
4000 1.38046886842267
4500 1.39768580963432
5000 1.40754530285057
};
\addplot [semithick, color2, mark=*, mark size=2, mark options={solid}]
table {%
0 1.47243335093012
500 1.48727239627895
1000 1.34464938415161
1500 1.43816114094896
2000 1.39259016241554
2500 1.24514126694347
3000 1.29912354586821
3500 1.32825716565459
4000 1.25320811051375
4500 1.20642675416557
5000 1.19469427678292
};
\addplot [semithick, color3, mark=triangle*, mark size=3, mark options={solid,rotate=180}]
table {%
0 1.47243335093012
500 1.18171057183751
1000 1.17290509812354
1500 1.12773421158836
2000 1.115865394892
2500 1.10560345221675
3000 1.10692321302297
3500 1.10491301203165
4000 1.09604383447748
4500 1.08248749436427
5000 1.0868835181893
};
\addplot [semithick, color4, mark=triangle*, mark size=3, mark options={solid}]
table {%
0 1.47243335093012
500 1.3183266640251
1000 1.25896733510687
1500 1.22247257465395
2000 1.19352159988236
2500 1.17271108041181
3000 1.15262003080406
3500 1.12559868873673
};
\end{groupplot}

\path ([yshift=+5mm]whatever c1r1.north) -- node[shift={(0,.7)}] {\ref{named}} ([yshift=+5mm]whatever c4r1.north);
\node[text width=6cm,align=center,anchor=south] at (whatever c1r1.north) {\subcaption{\label{fig:degree}}};
\node[text width=6cm,align=center,anchor=south] at (whatever c2r1.north) {\subcaption{\label{fig:diameter}}};
\node[text width=6cm,align=center,anchor=south] at (whatever c3r1.north) {\subcaption{\label{fig:successlong}}};
\node[text width=6cm,align=center,anchor=south] at (whatever c4r1.north) {\subcaption{\label{fig:feeslong}}};

\end{tikzpicture}
    }
    \caption{Long-term impact of attachment strategies on the network.}
    \label{fig:longterm}
\end{figure*}
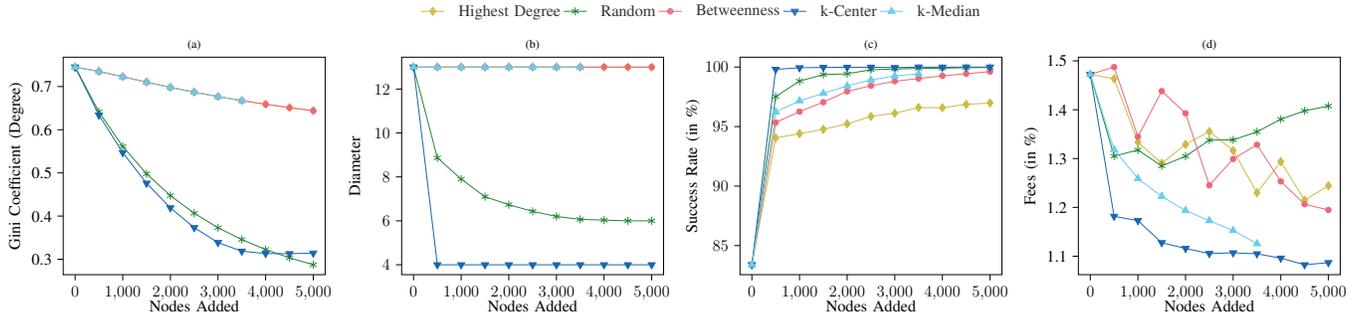
So far, we analyzed the discussed attachment strategies with respect to their
local short-term impact, \ie, from the point of view of an egoistical node joining the network.
In the following, we assess the global long-term impact of the discussed
attachment strategies for payment channel networks.

\subsection{Simulation Setup}
In order to evaluate the global long-term impact, we utilized the
time-discrete event-based network simulator from Section~\ref{sub:simulator} to model the
process of 5,000 nodes sequentially joining the Lightning Network, which
corresponds to more than doubling the network size. Each node
joins the network with $k=10$ channels that are established according to the
given strategy, which is roughly the network average node degree. While the
future network development will probably not exactly follow these
assumptions, this approach allows us to compare the advantages and
drawbacks of each strategy
without considering additionally interfering and confounding factors.
This simulation-based analysis was conducted for all but the \textsf{MBI} strategy. Due
to \textsf{MBI}'s significantly higher computational requirements
(cf.~Section~\ref{sec:runtimes}), we had to refrain from
including it in the long-term evaluation. As before, all randomized transactions were
repeated 1,000 and all simulations 30 times to ensure statistically
significant results.

\subsection{Impact on the Network's Topology}
In order to analyze the impact each attachment strategy has on network
centralization over time, we analyzed the network topology in intervals of
500 joining nodes and recorded essential network metrics.
Figure~\ref{fig:degree} shows the Gini coefficient of the node degree,
which quantifies the inequality of the degree distribution
for an increasing number of added nodes. As expected, the
network exhibits initially a high Gini value of nearly~$0.75$, which underlines the high degree of inequality currently exhibited by Lightning's
network topology. Furthermore, the results show that strategies following a preferential attachment pattern, such as the \textsf{Highest
Degree} or \textsf{Betweenness} strategies only marginally decrease the
centralization over time, while strategies
that also connect the fringes of the network, such as \textsf{$k$-Center} and
\textsf{Random} have a strong positive impact on centralization. Interestingly, we
observe that \textsf{$k$-Median} tends to elect the same set of $k$
nodes over time.
While these $k$ nodes increase their connectivity,
it does not result in a significant improvement
with respect to the degree inequality.

Figure~\ref{fig:diameter} shows the average network diameter,
\ie, the longest shortest path in the network, which is an indicator for
the worst-case routing complexity. Again, \textsf{Random} and
\textsf{$k$-Center} perform best and are able to immensely reduce the initial
network diameter of~13 already after attaching 500~nodes.
Notably, the \textsf{$k$-Center} strategy quickly allows all network nodes to
reach all other nodes in just four hops.
In order to get an understanding of how the participation in routing is impacted over
time, we analyzed the inequality of betweenness centralities and the central
point dominance. The results generally concur with our observations for node
degrees. They also show that our current choice of establishing the initial connection of the
\textsf{$k$-Center} and \textsf{$k$-Median} strategies to the single highest degree node
results in an increased central point dominance.
While this is an implementation detail, its impact requires further investigation in the future.

\subsection{Impact on the Network's Performance}
In order to evaluate the performance of the network in dependence of each
attachment strategy, we analyzed the average
success rate and the arising fees by regularly simulating transactions in the
network. To this end, we executed batches of 1,000 micro transactions with randomly chosen
sources and destinations after the addition of every 500 nodes.

As can be seen in Figure~\ref{fig:successlong}, the average network
success rate generally improves with an increasing number of nodes and the
additionally provided routing capacity. The evaluation moreover shows that
again the decentralizing strategies \textsf{Random} and especially
\textsf{$k$-Center} benefit the overall network connectivity the most, letting
the success rate quickly rise to close to 100\%.

While this pattern is generally also reflected in the average paid transaction
fees, as shown in Figure~\ref{fig:feeslong}, the results
highlight that a high degree of centralization can be beneficial for fee
costs. In particular, while the \textsf{Highest Degree} strategy does
generally not offer many benefits, it does result in rather low average fee
costs. This is likely due to the short average path lengths and high efficiency of star
sub-structures (cf.~\cite{avarikioti2020ride,sali2020optimizing}).
However, again the \textsf{$k$-Center} strategy proves to be the most
promising candidate to minimize fee costs for the end-user in the long term,
with \textsf{$k$-Median} being a close second.

\subsection{Discussion}
Throughout our analysis, it became apparent that the Lightning Network
currently is heavily restricted by its overall limited capacity
and its concentration on a few central service providers. We therefore found
that the provided quality of service and user experience would immensely
benefit from any kind of higher-volume and higher-connectivity adoption.

We also found that from an egoistical perspective, strategies selecting central attachment points seem to provide the best short-term performance, with
the exception of transaction fees, in which case the \textsf{$k$-Median}
strategy showed to be the most promising candidate. From the global
point of view, however, decentralizing strategies proved to
provide the best long-term benefits for the network overall.
With regard to this conflict of interest, we empirically confirm the trade-off between
efficiency and decentralization~\cite{waugh2020empirical,avarikioti2020ride}.

However, our analysis showed two strategies to be feasible and potentially
capable of combining local short-term and global long-term interests:
\textsf{$k$-Center} and \textsf{$k$-Median}.
While these strategies may not be the absolute optimum from the egoistical point
of view, they benefit the long-term network development the most. It therefore
remains an open question whether users would accept non-optimal short-term
strategies, if they benefit them and the whole network in the long-term. 

In order to balance this trade-off, real-world implementations should consider to employ a set of
different well-chosen strategies to establish their channels. However, the exact choices and the share
of connections established through a particular strategy are up to further
analysis. Our implementation of the \textsf{$k$-Center} and
\textsf{$k$-Median} strategies currently builds on an
initial centralized connection. We therefore also deem the potential of
such \enquote{mixed} strategies, \ie, strategies that further randomize and
distribute these connection types, a promising subject for future research.
Furthermore, while we generally hold the inherent conflicts of interest to be hard to
reconcile, we think they should further be discussed and addressed in the community.

\section{Related Work}\label{sec:relwork}
Most research on payment channel networks focuses on aspects such as the
channel design~\cite{hearn2015bitcoin,decker2015duplex,decker2018eltoo,lightning_paper,miller2017sprites}, the network's
topology~\cite{seres2019topology,beres2019simulator,rohrer2019dischargedPC}, and routing
algorithms~\cite{sivaraman2018spider,bagaria2020boomerang}.
While most of these entries take the network topology as a given, few entries
study how the network structure emerges and which algorithms for creation
are preferable. To this end, Avarikioti et al.~\cite{avarikioti2020ride} follow a game-theoretic approach
and show that centralized structures can make the network
more efficient and stable. In particular, the authors show that a star
graph, \ie, a graph with one central hub, poses a social optimum as well as a
Nash equilibrium in terms of efficiency and stability.
These results are supported by Sali and Zohar~\cite{sali2020optimizing}, who
show the efficiency of centralized hub structures.
Interestingly, Rincon et al.~\cite{rincon2020connectiontypes} come to the
conclusion that it is nonetheless not disadvantageous for smaller nodes, \ie,
nodes with few channels and a limited budget, to connect to other small nodes.
These connection types were even proven to positively impact the
network's robustness and efficiency, although connections to larger or richer
nodes were shown to improve the efficiency even further.

Most related to our work, Ersoy et al.~\cite{ersoy2019profit} study attachment
strategies and have been the first to observe that profit
maximization of service providers in the Lightning Network is connected to
their position in the network. To this end, the authors 
introduce the \emph{maximum reward improvement}~(MRI) problem, which, in
contrast to the MBI problem, additionally aims to optimize the joining node's fee policies.
While the authors reduce MRI to MBI and show that it is also NP-hard,
they provide an approximation algorithm.
The results underline that improving the betweenness
is a successful strategy to increase a node's fee revenue.
As the approximation algorithm proposed in~\cite{ersoy2019profit} is still
very costly, we deliberately refrain from
optimizing fee policies in our work.
Instead, we opt to implement a profit-oriented strategy
based on Bergamini et al.'s~\cite{bergamini2018betweenness} approximation algorithm under the assumption of default fee policies.

While many prior entries highlight that, in theory, attachment strategies optimizing for
profit and efficiency tend to favor the creation of highly centralized topologies, studies on the security~\cite{rohrer2019dischargedPC,mizrahi2020congestion,tochner2019hijacking,guo2019measurement}
and privacy~\cite{malavolta2017concurrency,malavolta2019anonymous,tikhomirov2020quantitative,kappos2020empirical,rohrer20countingdown}
of payment channel network emphasize the risk associated with network centralization.
These contradictory results indicate an inherent trade-off between
the network efficiency and decentralization, also observed by Waugh and
Holz~\cite{waugh2020empirical}. As this conflict of interest is not easily
resolved to one side or the other, additional insight on the long-term
effects of certain design decisions becomes necessary. To this end, we provide
the first empirical study on the long-term impact of attachment strategies for
payment channel networks.

Li et al.~\cite{balance_planning} present an algorithm that allows
to calculate the optimal distribution of initial channel balances assuming a
certain budget in order to satisfy payment demands. Similarly,
channel rebalancing protocols~\cite{revive,pickhardt2019rebalancing} aim to
optimally redistribute the allocation of funds in order to ensure frictionless
payment processing. We deem the integration of such capacity planning algorithms into
our model promising future work.

\section{Conclusion}\label{sec:conclusion}
In this work, we provided an empirical study on the impact of attachment strategies for payment
channel networks that once more exposes the fundamental trade-off between efficiency and
decentralization. While we were able to identify two candidate strategies with the
potential to combine local short-term and global long-term interests, we deem the question of attachment strategies an important
avenue for future research and discussion on payment channel networks.
\bibliographystyle{IEEEtran}
 \bibliography{master}

\begin{thebibliography}{10}
\providecommand{\url}[1]{#1}
\csname url@samestyle\endcsname
\providecommand{\newblock}{\relax}
\providecommand{\bibinfo}[2]{#2}
\providecommand{\BIBentrySTDinterwordspacing}{\spaceskip=0pt\relax}
\providecommand{\BIBentryALTinterwordstretchfactor}{4}
\providecommand{\BIBentryALTinterwordspacing}{\spaceskip=\fontdimen2\font plus
\BIBentryALTinterwordstretchfactor\fontdimen3\font minus
  \fontdimen4\font\relax}
\providecommand{\BIBforeignlanguage}[2]{{%
\expandafter\ifx\csname l@#1\endcsname\relax
\typeout{** WARNING: IEEEtran.bst: No hyphenation pattern has been}%
\typeout{** loaded for the language `#1'. Using the pattern for}%
\typeout{** the default language instead.}%
\else
\language=\csname l@#1\endcsname
\fi
#2}}
\providecommand{\BIBdecl}{\relax}
\BIBdecl

\bibitem{lightning_paper}
J.~Poon and T.~Dryja, ``The bitcoin lightning network: Scalable off-chain
  instant payments,'' 2016.

\bibitem{seres2019topology}
I.~A. Seres, L.~Guly{\'{a}}s, D.~A. Nagy, and P.~Burcsi, ``Topological analysis
  of bitcoin's lightning network,'' in \emph{Mathematical Research for
  Blockchain Economy}.\hskip 1em plus 0.5em minus 0.4em\relax Springer
  International Publishing, 2020, pp. 1--12.

\bibitem{lin2020centralisation}
\BIBentryALTinterwordspacing
J.~Lin, K.~Primicerio, T.~Squartini, C.~Decker, and C.~J. Tessone, ``Lightning
  network: a second path towards centralisation of the bitcoin economy,''
  \emph{CoRR}, vol. abs/2002.02819, 2020. [Online]. Available:
  \url{https://arxiv.org/abs/2002.02819}
\BIBentrySTDinterwordspacing

\bibitem{rohrer2019dischargedPC}
E.~Rohrer, J.~Malliaris, and F.~Tschorsch, ``Discharged payment channels:
  Quantifying the lightning network's resilience to topology-based attacks,''
  in \emph{2019 {IEEE} European Symposium on Security and Privacy Workshops,
  EuroS{\&}P Workshops 2019, Stockholm, Sweden, June 17-19, 2019}.\hskip 1em
  plus 0.5em minus 0.4em\relax {IEEE}, 2019, pp. 347--356.

\bibitem{tochner2019hijacking}
\BIBentryALTinterwordspacing
S.~Tochner, S.~Schmid, and A.~Zohar, ``Hijacking routes in payment channel
  networks: {A} predictability tradeoff,'' \emph{CoRR}, vol. abs/1909.06890,
  2019. [Online]. Available: \url{http://arxiv.org/abs/1909.06890}
\BIBentrySTDinterwordspacing

\bibitem{kappos2020empirical}
G.~Kappos, H.~Yousaf, A.~Piotrowska, S.~Kanjalkar, S.~Delgado-Segura,
  A.~Miller, and S.~Meiklejohn, ``An empirical analysis of privacy in the
  lightning network,'' \emph{arXiv preprint arXiv:2003.12470}, 2020.

\bibitem{rohrer20countingdown}
E.~Rohrer and F.~Tschorsch, ``Counting down thunder: Timing attacks on privacy
  in payment channel networks,'' in \emph{AFT '20: Proceedings of the second
  ACM conference on Advances in Financial Technologies}.

\bibitem{waugh2020empirical}
\BIBentryALTinterwordspacing
F.~Waugh and R.~Holz, ``An empirical study of availability and reliability
  properties of the bitcoin lightning network,'' \emph{CoRR}, vol.
  abs/2006.14358, 2020. [Online]. Available:
  \url{https://arxiv.org/abs/2006.14358}
\BIBentrySTDinterwordspacing

\bibitem{avarikioti2020ride}
Z.~Avarikioti, L.~Heimbach, Y.~Wang, and R.~Wattenhofer, ``Ride the lightning:
  The game theory of payment channels,'' in \emph{{FC}~'20: Proceedings of the
  24th International Conference on Financial Cryptography and Data Security},
  Feb. 2020, pp. 264--283.

\bibitem{sali2020optimizing}
\BIBentryALTinterwordspacing
Y.~Sali and A.~Zohar, ``Optimizing off-chain payment networks in
  cryptocurrencies,'' \emph{CoRR}, vol. abs/2007.09410, 2020. [Online].
  Available: \url{https://arxiv.org/abs/2007.09410}
\BIBentrySTDinterwordspacing

\bibitem{ersoy2019profit}
O.~Ersoy, S.~Roos, and Z.~Erkin, ``How to profit from payments channels,'' in
  \emph{Financial Cryptography and Data Security - 24th International
  Conference, {FC} 2020, Kota Kinabalu, Malaysia, February 10-14, 2020 Revised
  Selected Papers}, ser. Lecture Notes in Computer Science, vol. 12059.\hskip
  1em plus 0.5em minus 0.4em\relax Springer, 2020, pp. 284--303.

\bibitem{hearn2015bitcoin}
\BIBentryALTinterwordspacing
M.~Hearn and J.~Spilman. (2015) Bitcoin contracts. [Online]. Available:
  \url{https://en.bitcoin.it/wiki/Contracts}
\BIBentrySTDinterwordspacing

\bibitem{decker2015duplex}
C.~Decker and R.~Wattenhofer, ``A fast and scalable payment network with
  bitcoin duplex micropayment channels,'' in \emph{Stabilization, Safety, and
  Security of Distributed Systems - 17th International Symposium, {SSS} 2015,
  Edmonton, AB, Canada, August 18-21, 2015, Proceedings}, ser. Lecture Notes in
  Computer Science, vol. 9212.\hskip 1em plus 0.5em minus 0.4em\relax Springer,
  2015, pp. 3--18.

\bibitem{miller2017sprites}
\BIBentryALTinterwordspacing
A.~Miller, I.~Bentov, R.~Kumaresan, and P.~McCorry, ``Sprites: Payment channels
  that go faster than lightning,'' \emph{CoRR}, vol. abs/1702.05812, 2017.
  [Online]. Available: \url{http://arxiv.org/abs/1702.05812}
\BIBentrySTDinterwordspacing

\bibitem{decker2018eltoo}
C.~Decker, R.~Russell, and O.~Osuntokun, ``eltoo: A simple layer2 protocol for
  bitcoin,'' \emph{White paper: https://blockstream.com/eltoo.pdf}, 2018.

\bibitem{snapshots}
\BIBentryALTinterwordspacing
E.~Rohrer. Snapshots of the lightning network. [Online]. Available:
  \url{https://gitlab.tubit.tu-berlin.de/rohrer/discharged-pc-data/tree/master/snapshots}
\BIBentrySTDinterwordspacing

\bibitem{mccorry2016towards}
P.~McCorry, M.~M{\"{o}}ser, S.~F. Shahandashti, and F.~Hao, ``Towards bitcoin
  payment networks,'' in \emph{Information Security and Privacy - 21st
  Australasian Conference, {ACISP} 2016, Melbourne, VIC, Australia, July 4-6,
  2016, Proceedings, Part {I}}, ser. Lecture Notes in Computer Science, vol.
  9722.\hskip 1em plus 0.5em minus 0.4em\relax Springer, 2016, pp. 57--76.

\bibitem{danezis2009sphinx}
G.~Danezis and I.~Goldberg, ``Sphinx: {A} compact and provably secure mix
  format,'' in \emph{SP~'09: Proceedings of the 30th IEEE Symposium on Security
  and Privacy}, 2009, pp. 269--282.

\bibitem{dijkstra1959note}
E.~W. Dijkstra, ``A note on two problems in connexion with graphs,''
  \emph{Numerische Mathematik}, vol.~1, pp. 269--271, 1959.

\bibitem{harris2020flood}
\BIBentryALTinterwordspacing
J.~Harris and A.~Zohar, ``Flood {\&} loot: {A} systemic attack on the lightning
  network,'' \emph{CoRR}, vol. abs/2006.08513, 2020. [Online]. Available:
  \url{https://arxiv.org/abs/2006.08513}
\BIBentrySTDinterwordspacing

\bibitem{barabasi1999emergence}
A.-L. Barab{\'a}si and R.~Albert, ``Emergence of scaling in random networks,''
  \emph{science}, vol. 286, no. 5439, pp. 509--512, 1999.

\bibitem{lnd_autopilot_issue}
\BIBentryALTinterwordspacing
R.~Pickhardt. Is the barabási-albert model a reasonable choice for the
  autopilot? [Online]. Available:
  \url{https://github.com/lightningnetwork/lnd/issues/677}
\BIBentrySTDinterwordspacing

\bibitem{freeman1977betweenness}
L.~C. Freeman, ``A set of measures of centrality based on betweenness,''
  \emph{Sociometry}, pp. 35--41, 1977.

\bibitem{Brandes08}
U.~Brandes, ``On variants of shortest-path betweenness centrality and their
  generic computation,'' \emph{Social Networks}, vol.~30, no.~2, pp. 136--145,
  2008.

\bibitem{fastBC}
M.~Baglioni, F.~Geraci, M.~Pellegrini, and E.~Lastres, ``Fast exact computation
  of betweenness centrality in social networks,'' in \emph{International
  Conference on Advances in Social Networks Analysis and Mining, {ASONAM} 2012,
  Istanbul, Turkey, 26-29 August 2012}.\hskip 1em plus 0.5em minus 0.4em\relax
  {IEEE} Computer Society, 2012, pp. 450--456.

\bibitem{hochbaum1985k_center}
D.~S. Hochbaum and D.~B. Shmoys, ``A best possible heuristic for the
  \emph{k}-center problem,'' \emph{Mathematics of Operations Research},
  vol.~10, no.~2, pp. 180--184, 1985.

\bibitem{gonzalez}
T.~F. Gonzalez, ``Clustering to minimize the maximum intercluster distance,''
  \emph{Theoretical Computer Science}, vol.~38, pp. 293--306, 1985.

\bibitem{meyerson2009min_avgSP}
A.~Meyerson and B.~Tagiku, ``Minimizing average shortest path distances via
  shortcut edge addition,'' in \emph{Approximation, Randomization, and
  Combinatorial Optimization. Algorithms and Techniques, 12th International
  Workshop, {APPROX} 2009, and 13th International Workshop, {RANDOM} 2009,
  Berkeley, CA, USA, August 21-23, 2009. Proceedings}, ser. Lecture Notes in
  Computer Science, vol. 5687.\hskip 1em plus 0.5em minus 0.4em\relax Springer,
  2009, pp. 272--285.

\bibitem{chrobak2005RGreedy}
M.~Chrobak, C.~Kenyon, and N.~E. Young, ``The reverse greedy algorithm for the
  metric \emph{K}-median problem,'' in \emph{Computing and Combinatorics, 11th
  Annual International Conference, {COCOON} 2005, Kunming, China, August 16-29,
  2005, Proceedings}, ser. Lecture Notes in Computer Science, vol. 3595.\hskip
  1em plus 0.5em minus 0.4em\relax Springer, 2005, pp. 654--660.

\bibitem{bergamini2018betweenness}
E.~Bergamini, P.~Crescenzi, G.~D'Angelo, H.~Meyerhenke, L.~Severini, and
  Y.~Velaj, ``Improving the betweenness centrality of a node by adding links,''
  \emph{{ACM} Journal of Experimental Algorithmics}, vol.~23, 2018.

\bibitem{beres2019simulator}
\BIBentryALTinterwordspacing
F.~B{\'{e}}res, I.~A. Seres, and A.~A. Bencz{\'{u}}r, ``A cryptoeconomic
  traffic analysis of bitcoins lightning network,'' \emph{CoRR}, vol.
  abs/1911.09432, 2019. [Online]. Available:
  \url{http://arxiv.org/abs/1911.09432}
\BIBentrySTDinterwordspacing

\bibitem{sivaraman2018spider}
V.~Sivaraman, S.~B. Venkatakrishnan, M.~Alizadeh, G.~C. Fanti, and
  P.~Viswanath, ``Routing cryptocurrency with the spider network,'' in
  \emph{HotNets~'18: Proceedings of the 17th {ACM} Workshop on Hot Topics in
  Networks}, Nov. 2018, pp. 29--35.

\bibitem{bagaria2020boomerang}
V.~K. Bagaria, J.~Neu, and D.~Tse, ``Boomerang: Redundancy improves latency and
  throughput in payment-channel networks,'' in \emph{{FC}~'20: Proceedings of
  the 24th International Conference on Financial Cryptography and Data
  Security}, Feb. 2020, pp. 304--324.

\bibitem{rincon2020connectiontypes}
D.~Rincon, E.~Y. Wu, S.~Dewar, and D.~Zhu, ``Identifying beneficial connection
  types in payment channel networks: The case of lightning,'' \emph{University
  of California Berkeley}, 2020.

\bibitem{mizrahi2020congestion}
A.~Mizrahi and A.~Zohar, ``Congestion attacks in payment channel networks,''
  \emph{CoRR}, vol. abs/2002.06564, 2020.

\bibitem{guo2019measurement}
Y.~Guo, J.~Tong, and C.~Feng, ``A measurement study of bitcoin lightning
  network,'' in \emph{{IEEE} International Conference on Blockchain, Blockchain
  2019, Atlanta, GA, USA, July 14-17, 2019}.\hskip 1em plus 0.5em minus
  0.4em\relax {IEEE}, 2019, pp. 202--211.

\bibitem{malavolta2017concurrency}
G.~Malavolta, P.~Moreno{-}Sanchez, A.~Kate, M.~Maffei, and S.~Ravi,
  ``Concurrency and privacy with payment-channel networks,'' in \emph{CCS~'17:
  Proceedings of the 2017 {ACM} {SIGSAC} Conference on Computer and
  Communications Security}, Oct. 2017, pp. 455--471.

\bibitem{malavolta2019anonymous}
G.~Malavolta, P.~Moreno{-}Sanchez, C.~Schneidewind, A.~Kate, and M.~Maffei,
  ``Anonymous multi-hop locks for blockchain scalability and
  interoperability,'' in \emph{NDSS~'19: Prooceedings of the 26th Annual
  Network and Distributed System Security Symposium}, Feb. 2019.

\bibitem{tikhomirov2020quantitative}
S.~Tikhomirov, P.~Moreno{-}Sanchez, and M.~Maffei, ``A quantitative analysis of
  security, anonymity and scalability for the lightning network,'' \emph{{IACR}
  Cryptol. ePrint Arch.}, vol. 2020, p. 303, 2020.

\bibitem{balance_planning}
P.~Li, T.~Miyazaki, and W.~Zhou, ``Secure balance planning of off-blockchain
  payment channel networks,'' 2020.

\bibitem{revive}
R.~Khalil and A.~Gervais, ``Revive: Rebalancing off-blockchain payment
  networks,'' in \emph{Proceedings of the 2017 {ACM} {SIGSAC} Conference on
  Computer and Communications Security, {CCS} 2017, Dallas, TX, USA, October 30
  - November 03, 2017}.\hskip 1em plus 0.5em minus 0.4em\relax {ACM}, 2017, pp.
  439--453.

\bibitem{pickhardt2019rebalancing}
\BIBentryALTinterwordspacing
R.~Pickhardt and M.~Nowostawski, ``Imbalance measure and proactive channel
  rebalancing algorithm for the lightning network,'' \emph{CoRR}, vol.
  abs/1912.09555, 2019. [Online]. Available:
  \url{http://arxiv.org/abs/1912.09555}
\BIBentrySTDinterwordspacing

\bibitem{fc20}
\emph{{FC}~'20: Proceedings of the 24th International Conference on Financial
  Cryptography and Data Security}, Feb. 2020.

\end{thebibliography}

\end{document}